\documentclass[twocolumn,secnumarabic,amssymb, nobibnotes, aps, prd]{revtex4-2}
\usepackage{amssymb}
\usepackage{graphicx}
\usepackage{amsmath}
\usepackage{caption}
\usepackage{subcaption}
\usepackage{multirow}

\begin{document}

\title{Influence of chemical potential to explain the maximum mass and tidal love number of strange stars}%

\author{Debadri Bhattacharjee \& Pradip Kumar Chattopadhyay} %% Author name

\address{IUCAA Centre for Astronomy Research and Development (ICARD), Department of Physics, Cooch Behar Panchanan Barma University, Vivekananda Street, District: Cooch Behar,  Pin: 736101, West Bengal, India.}
\email{debadriwork@gmail.com \& pkc\_76@rediffmail.com}
\begin{abstract}
	We investigate the influence of medium effects on strange quark matter and their consequences for the structural properties of compact stars. In this study, the bag constant, in the MIT bag model equation of state, is reformulated as a function of the chemical potential of quark. Imposing the Bodmer-Witten stability criterion, the parameter space is constrained by evaluating the energy per baryon. The Tolman-Oppenheimer-Volkoff equations are then solved to obtain the maximum mass-radius configurations. Our analysis shows that an increase in chemical potential leads to a reduction in the effective value of bag constant, resulting in a stiffer equation of state and correspondingly higher value of maximum stellar mass. Furthermore, we examine the chemical potential dependence of the tidal Love number and tidal deformability along with their corresponding finite surface density corrections. The results demonstrate that, for the chosen set of parameters, the constraint from the GW170817 event, namely $\Lambda < 800$, is consistently satisfied. In this paper, we have tried to establish how chemical potential $(\mu)$ affects the maximum mass and tidal deformability of strange stars.
\end{abstract}

\maketitle
%\tableofcontents

\section{Introduction}\label{sec1}
Neutron stars (NS) were first hypothesised, as remnants of supernovae, by W. Baade and F. Zwicky \cite{Baade}. Later, the first discovery of pulsars \cite{Jocelyn} presented the decisive evidence of NS. Since then the investigations of the interior composition NS has become one of the pivotal points in astrophysics. Over the years, several theories and notions have been presented to simulate the internal matter distribution. Among them, the Quark star hypothesis has stood as one of the foundational cornerstones. Recent advancements in observational astronomy have significantly improved our ability to study high-energy astrophysical phenomena on compact scales, leading to renewed interest in understanding the behaviour of matter under the extreme conditions of high densities and low temperatures encountered in NS. Since replicating such extreme conditions in terrestrial laboratories is currently unfeasible, the equation of state (EoS) for ultra-dense matter primarily depends on theoretical modeling \cite{Baym,Feroci}. Considerable research efforts have been devoted to this problem, particularly in probing the microscopic composition of compact stars. Various exotic states of matter have been proposed in this context, such as high-temperature superconducting phases, quark-gluon plasma, and Bose-Einstein condensates.

For NS and quark stars, formulating a reliable EoS for strongly interacting matter at supranuclear densities and low temperatures is essential. In principle, such an EoS can be derived from Quantum Chromodynamics (QCD), the fundamental theory governing strong interactions. However, QCD becomes highly non-perturbative in this regime, and lattice QCD computations face severe limitations due to the so-called "sign problem" at finite baryon chemical potential \cite{Forcrand}. At lower densities, where nucleonic degrees of freedom dominate, effective field theories-both relativistic and non-relativistic-are employed to describe nuclear interactions \cite{Lattimer}.

In the inner cores of compact stars, where the density exceeds nuclear saturation density by several times, it becomes a necessity to consider models based on deconfined quark matter. The concept that quark matter may exist in the core of NS was first proposed by Itoh \cite{Itoh}. Subsequent influential studies, especially by Madsen \cite{Madsen}, highlighted the role of strange quarks in enhancing the stability of compact objects, thereby giving rise to the notion of Strange Stars (SS) or Strange Quark Stars (SQS) \cite{Madsen,Baym1,Alcock,Glendenning}. These objects are hypothesised to be composed entirely of Strange Quark Matter (SQM), which has been conjectured by Witten \cite{Witten} to be the true ground state of QCD. 

Owing to the absence of fully realistic treatments for this high-density regime, simplified phenomenological models are often used. One such model is the MIT bag model \cite{Chodos} where the bag parameter is included in the thermodynamic potential of a free fermionic system to effectively describe the confinement of quarks within a finite region. The bag parameter effectively describes the energy differences between the perturbative and non-perturbative vacua \cite{Burgio,Burgio1} and usually for a minimalistic approach, it is taken to be constant. Even though MIT bag model contains an underlying assumption which presumes that quarks behave as free particles inside the bag, the quark matter does not achieve the asymptotic freedom instantaneously during or immediately after the phase transition. To tackle this situation, medium effects in quark star model are introduced. In this context, Farhi and Jaffe \cite{Farhi} explored the impact of finite strange quark mass on the properties of SQM. Later, Chakrabarty et al. \cite{Chakrabarty} investigated the quark confinement mechanism through quark mass dependent model. A comparable effect can be modeled within the bag framework by introducing a bag pressure that varies with density. It is widely recognised that, at the high baryon densities pertinent to the interiors of NS, quarks tend to exhibit asymptotic freedom \cite{Burgio,Burgio1}. This behaviour supports the adoption of a density-dependent bag pressure instead of assuming a constant value. Several works are available in literature that focus on the density-dependent approach \cite{Sen,PKC,KBG,Pal1}. 

The MIT bag model \cite{Chodos} has long served as one of the most prominent frameworks for describing the interior composition of SS. Over the years, numerous investigations have been devoted to exploring its implications, particularly under the assumption of a constant bag parameter $(B_{g})$ \cite{Kettner}. As demonstrated by \cite{Madsen}, the permissible range of $B_{g}$ is tightly constrained by the requirement of strange matter stability. The lower bound, $B_{g}^{1/4}=145~\mathrm{MeV}$ or $B_{g}=57.55~\mathrm{MeV/fm^3}$, marks the threshold below which strange quark matter becomes unstable. In systems containing only up $(u)$ and down $(d)$ quarks, the configuration is energetically unfavourable, as also discussed in \cite{Madsen}. The incorporation of strange $(s)$ quarks, however, reduces the energy per baryon and enhances overall stability. For non-strange (two-flavour) quark matter to remain stable relative to nuclear matter, its energy per baryon must be less than the neutron mass, $939.6~\mathrm{MeV}$ \cite{Madsen}, at zero external pressure. This requirement determines the minimum $B_{g}$ value necessary to prevent ordinary nuclei from decaying into deconfined quark matter. Conversely, the upper limit on $B_{g}$ follows from the condition that strange quark matter must be energetically more stable than the most tightly bound nuclei, such as iron. This criterion yields $(B_{g})^{1/4}_\mathrm{{max}}= 162.8~\mathrm{MeV}$ or equivalently $(B_{g})_\mathrm{{max}}=91.54~\mathrm{MeV/fm^3}$ \cite{Madsen}. When the comparison is instead made with respect to a neutron at zero pressure, the upper bound slightly increases to $(B_{g})^{1/4}_{\mathrm{max}}=164.4~\mathrm{MeV}$, corresponding to $(B_{g})_\mathrm{{max}}=95.11~\mathrm{MeV/fm^3}$ \cite{Kapusta,Madsen}.

The energy per baryon, $E_{B}$ of most stable nuclei, $^{56}Fe$, is $E_{B}=\frac{\rho}{n}=930.4~MeV$ \cite{Madsen}. Following the Bodmer-Witten hypothesis \cite{Bodmer,Witten}, the absolute stability of SQM is evaluated by calculating $E_{B}$ at zero external pressure, relative to neutron, and if $E_{B}<930.4~MeV$, SQM is termed stable. Now, from the thermodynamic relation for pressure, i.e., $p=\rho^{2}\frac{d}{d\rho}(\frac{\rho}{n})$, it must be noted that the minimum of $E_{B}$ must occur at zero external pressure. However, a reformulated quark mass dependent model \cite{Benvenuto,Lugones} shows inconsistency to that conjecture. A recent investigation \cite{Lugones1} into SQM within the framework of a density dependent quark mass model addresses and solves this internal inconsistencies by employing the canonical ensemble, and within this formalism, $E_{B}$ approaches a minimum value at zero pressure and the EoS is formulated via Euler formalism in the form, 
\begin{equation}
	\rho=-p+\sum_{i}\mu_{i}n_{i}, \label{eq1}
\end{equation}
where $\rho$ is the energy density, $p$ is the pressure, $\mu_{i}$ and $n_{i}$ represent the chemical potential and number density of the $i^{th}$ particle.

In general, the grand canonical ensemble formalism is employed to derive the EoS for SQM. However, incorporating a density dependent bag within this framework leads to thermodynamic inconsistency, as it violates the Euler relation. To address this issue, we advocate for the adoption of a chemical potential dependent bag model $B(\mu)$ in place of the density dependent form, ensuring consistency within the framework of grand canonical ensemble. The primary objective of this work is to employ a chemical potential-dependent bag model to construct compact stellar configurations. We aim to investigate the resulting mass-radius relations and tidal deformability characteristics of various compact object candidates, with a particular focus on the accuracy in describing the low-mass compact stars. 

The rest of the paper is organised in the following way: the formalism related to chemical potential dependent MIT bag model is described in Section~\ref{sec2}. In addition, Section~\ref{sec2} provides a detailed thermodynamic analysis showing why chemical potential dependent models must be favoured over density dependent models in the grand canonical ensemble. In Section~\ref{sec3}, we have addressed the necessary bounds on bag parameter $(B(\mu))$, and chemical potential $(\mu)$, obtained by utilising the the $E_{B}-\mu$ plot. The maximum mas-radius relations are obtained in Section~\ref{sec4}. Section~\ref{sec5} demonstrates the variations of tidal parameters, namely, tidal deformability, and tidal love number in the present scenario. Finally, we conclude the main findings in Section~\ref{sec6}.   

\section{Density vs chemical potential dependent bag: Thermodynamic analysis in grand canonical ensemble}\label{sec2}
\begin{figure}[h!]
	\centering
	\includegraphics[width=8cm]{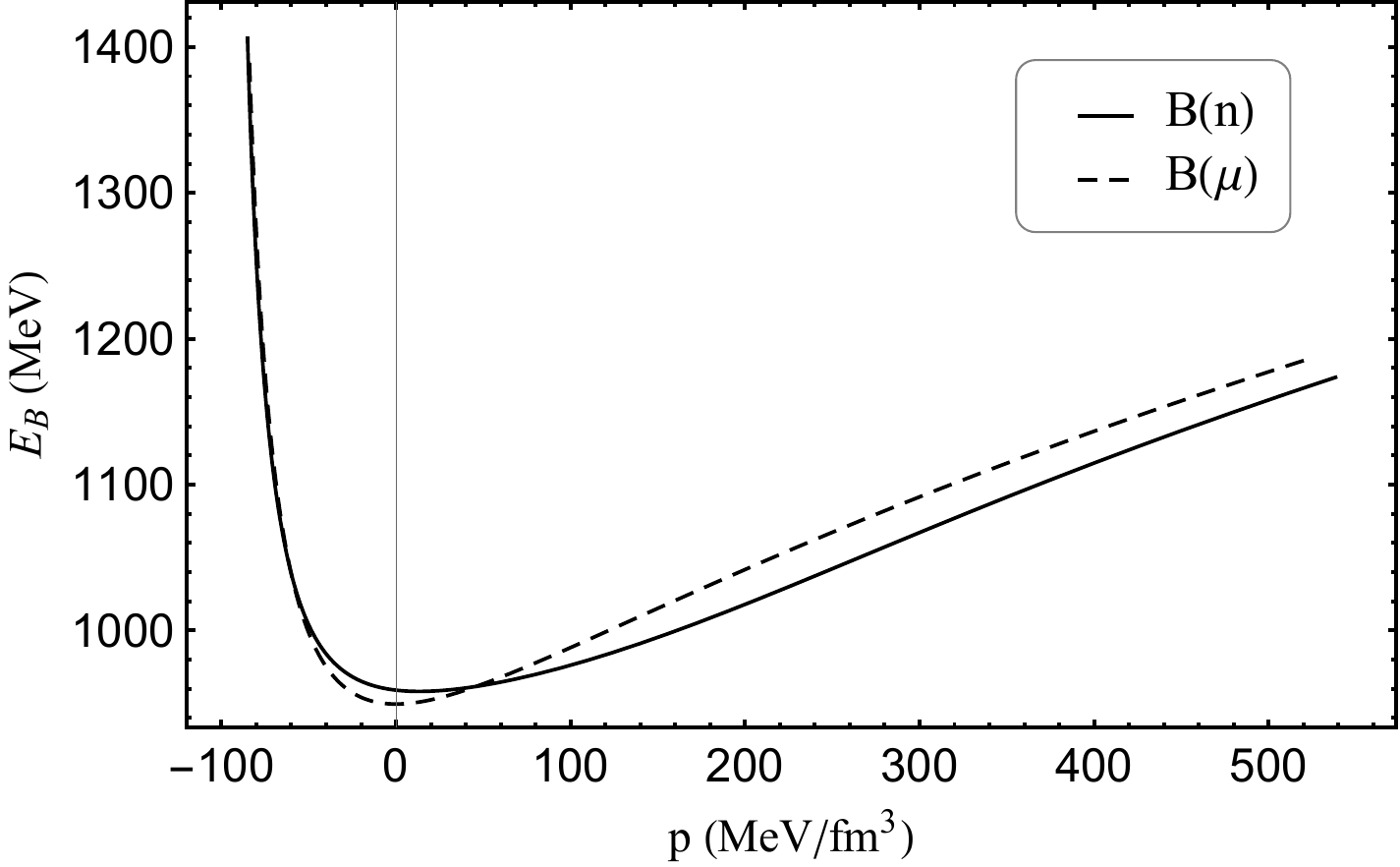}
	\caption{$E_{B}$ vs. $p$ for density dependent and chemical potential dependent approaches. The parameter choices for the figure are as follows: $B(n)$ : $\beta_{n}=0.1$, $n_{0}=0.152~fm^{-3}$, $B_{0}=100~MeV/fm^{3}$ and $B_{as}=80~MeV/fm^{3}$. $B(\mu)$ : $\beta_{\mu}=0.8$, $\mu_{0}=1000~fm^{-3}$, $B_{0}=100~MeV/fm^{3}$ and $B_{as}=30~MeV/fm^{3}$.}
	\label{fig1a}
\end{figure}

In this section, within the framework of grand canonical ensemble, we provide a thermodynamic comparison between density and chemical potential dependent bag models. We start by adopting the grand canonical potential, $\Omega~(T,V,\mu)$ that depends on temperature $(T)$, volume $(V)$, and chemical potential $(\mu)$, and all the thermodynamic parameters can be derived from $\Omega$. Further, we restrict ourselves to the cold compact star model, implying $T\rightarrow0$. In the grand canonical ensemble, the pressure is expressed as:
\begin{equation}
	p=-\left(\frac{d\Omega}{dV}\right)_{T,\mu}. \label{eq1a}
\end{equation}
\begin{itemize}
	\item Density dependent bag model: Let us consider, $\Omega_{g}=\frac{\Omega}{V}$ is the grand canonical potential density, and the number density, $n=\frac{N}{V}$. Substituting these in Eq.~(\ref{eq1a}), we obtain:
	\begin{equation}
		p=-\left[\frac{d\,(\Omega_{g}V)}{d\left(\frac{N}{n}\right)}\right]_{T,\mu}=-\Omega_{g}+n\left(\frac{\partial\Omega_{g}}{\partial n}\right)_{T,\mu}. \label{eq1b}
	\end{equation}
	Notably, the second term in the right hand side of Eq.~(\ref{eq1b}) appears due to the density dependency. Now, in the context of quark matter, $\Omega_{g}$ is written as:
	\begin{equation}
		\Omega_{g}=\sum_{f=u,d,s}\Omega_{f}+B(n), \label{eq1c}
	\end{equation}
	where $\Omega_{f}=-\frac{1}{4\pi^{2}}\Bigg[\mu_{f}\sqrt{\mu_{f}^{2}-m_{f}^{2}}\left(\mu_{f}^2-\frac{5}{2}m_{f}^{2}\right)+\frac{3}{2}m_{f}^{4}$ $ln\left[\frac{\mu_{f}+\sqrt{\mu_{f}^{2}-m_{f}^{2}}}{m_{f}}\right]\Bigg]$, $\mu_{f}$ and $m_{f}$, respectively, represent the chemical potential and mass of the free quark matter and the medium effects are incorporated by the bag, $B(n)$. Therefore, in this framework, the pressure is reformulated as:
	\begin{equation}
		p=-\Omega_{g}+n\frac{\partial B(n)}{\partial n}. \label{eq1d}
	\end{equation}
	Considering, Eq.~(\ref{eq1c}), $\sum_{f}m_{f}=0$ and a system of 3-flavour quarks, we can further simplify Eq.~(\ref{eq1d}) in the form:
	\begin{equation}
		p=\frac{3\mu^{4}}{4\pi^{2}}-B(n)+n\frac{\partial B(n)}{\partial n}. \label{eq1dd}
	\end{equation}
	The number density of quarks can be completely expressed as:
	\begin{equation}
		n=\frac{1}{3}\sum_{f}n_{f}=-\frac{1}{3}\sum_{f}\frac{\partial\Omega_{f}}{\partial\mu_{f}}=\sum_{f}\left(\frac{k_{f}^{3}}{\pi^{2}}-\frac{\partial B(\mu)}{\partial\mu_{f}}\right), \label{eq1g}
	\end{equation} 
	where, $k_{f}=\sqrt{\mu_{f}^{2}-m_{f}^{2}}$ denotes the Fermi momentum of the free quark matter. In the $n$-dependent approach, the number density takes the form:
	\begin{equation}
		n=\sum_{f}\left(\frac{k_{f}^{3}}{\pi^{2}}\right). \label{eq1ggg}
	\end{equation}
	Using Eq.~(\ref{eq1}), we obtain the energy density in the form:
	\begin{equation}
		\rho=-p+\frac{3\mu^{4}}{\pi^{2}}. \label{eq1ee}
	\end{equation}
	It must be noted that in the density dependent approach, the last term on the right hand side of Eq.~(\ref{eq1g}) drops out due to its $\mu$-dependence. Furthermore, from Eq.~(\ref{eq1}), it must be noted that the Euler relation depends on chemical potential $(\mu)$. So, Eq.~(\ref{eq1d}) should also depend on $\mu$, instead of $n$. Hence, the Euler relation is violated in this approach.   
	\item Chemical potential dependent bag model: To resolve the inconsistency, we consider a $\mu$ dependent bag, $B(\mu)$ \cite{Pal}. Now, following the same steps as Eqs.~(\ref{eq1a}) and (\ref{eq1b}), we obtain:
	\begin{equation}
		p=-\Omega_{g}, \label{eq1e}
	\end{equation}
	and 
	\begin{equation}
		\Omega_{g}=\sum_{f=u,d,s}\Omega_{f}+B(\mu), \label{eq1f}
	\end{equation} 
	Following Eqs.~(\ref{eq1}), (\ref{eq1g}) and (\ref{eq1f}) along with the condition that $\sum_{f}m_{f}=0$, we obtain:
	\begin{equation}
		p=\frac{3\mu^{4}}{4\pi^{2}}-B(\mu), \label{eq1h}
	\end{equation}
	\begin{equation}
		n=\frac{\mu^{3}}{\pi^{2}}-\frac{1}{3}\frac{\partial B(\mu)}{\partial\mu}, \label{eq1i}
	\end{equation}
	and 
	\begin{equation}
		\rho=-p+\frac{3\mu^{4}}{\pi^{2}}-\mu\frac{\partial B(\mu)}{\partial\mu}. \label{eq1j}
	\end{equation}
	Hence, it is evident that chemical potential dependent bag model respects the Euler relation and provides a complete framework to study quark matter in grand canonical ensemble.	
\end{itemize}
Furthermore, to clarify the thermodynamic consistency, we have considered the following Gaussian parametrisations of bag for both density and chemical potential dependent approaches: 
\begin{itemize}
	\item Following the work of Burgio et al. \cite{Burgio}, we have considered the following form:
	\begin{equation}
		B(n)=B_{as}+(B_{0}-B_{as})e^{-\beta_{n}(\frac{n}{n_{0}})^2}, \label{eq1ff}
	\end{equation}
	where, $B_{as}$ represents the finite value of $B$ at asymptotic densities, $B_{0}=B(\rho=0)$, $\beta_{n}$ is a free parameter, $n$ is the baryon number density and $n_{0}$ represents the baryon number density of ordinary nuclear matter.
	\item Following the work of Pal and Chaudhuri \cite{Pal}, we consider the Gaussian expression of bag parameter $(B(\mu))$, in the following form:
	\begin{equation}
		B(\mu)=B_{as}+(B_{0}-B_{as})e^{-\beta_{\mu}(\frac{\mu}{\mu_{0}})^{2}}, \label{eq2}
	\end{equation}
	where, $B_{as}$ denotes the asymptotic value of $B(\mu)$, $B_{0}=B(\mu=0)$, $\beta_{\mu}$ denotes a control parameter and $\mu_{0}$ represents the saturation chemical potential.
\end{itemize}
Using the forms given in Eqs.~(\ref{eq1ff}) and (\ref{eq2}), we have demonstrated the variation of energy per baryon $(E_{B}=\frac{\rho}{n})$ with pressure $(p)$, for different values of $\mu$ \cite{Pal}, in Fig.~(\ref{fig1a}). $\rho$ is taken from Eqs.~(\ref{eq1ee}) and (\ref{eq1j}) and $n$ is taken from Eqs.~(\ref{eq1ggg}) and (\ref{eq1i}).

From Fig.~(\ref{fig1a}) we note that the minimum energy per baryon occurs at zero pressure for the chemical potential dependent bag. Moreover, the chemical potential dependent approach maintains the Bodmer-Witten hypothesis \cite{Bodmer} pertaining to the stable SQM, i.e., $E_{B}<930.4~MeV$ at zero external pressure. As the pressure is not zero inside the star, $E_{B}$ may exceed 930.4 MeV, without violating the Bodmer-Witten hypothesis \cite{Witten,Bodmer}. So far the modeling of stable bare strange stars is concerned, the value of $E_{B}<930.4~MeV$ must hold at the surface at zero pressure. 

From Eq.~(\ref{eq1j}), it must be noted that the density of each quark flavour is altered due to the bag parameter, which now depends on the chemical potential. Now, following the framework introduced by Kettner et al. \cite{Kettner}, we may now modify the MIT bag model in presence of $B(\mu)$ as:
\begin{equation}
	p=\frac{1}{3}\Big(\rho-4B(\mu)\Big). \label{eq6}
\end{equation}
In the present paper, we have used this EoS to study the physical features of compact objects. 
\section{Bounds on $B(\mu)$ from $E_{B}-\mu$ plot}\label{sec3}
\begin{figure}[h!]
	\centering
	\includegraphics[width=8cm]{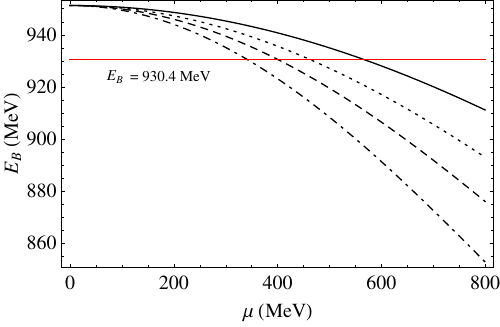}
	\caption{$E_{B}$ vs. $\mu$ Plot for different $\beta_{\mu}$. Here, the solid, dotted, dashed and dotdashed lines represent, respectively, $\beta_{\mu}=0.4,~0.6,~0.8$ and $1.10$.}
	\label{fig1}
\end{figure}

The parameters associated with the Gaussian form of the bag function, $B(\mu)$, are adopted from the work of Pal and Chaudhuri \cite{Pal}. In their analysis, it is demonstrated that the stability of strange quark matter (SQM) is governed by the energy per baryon, in accordance with the Bodmer-Witten hypothesis \cite{Witten,Bodmer}. Consequently, the allowed ranges of the model parameters must be constrained so as to satisfy this hypothesis. Following the Bodmer-Witten criterion and the results presented in Ref.~\cite{Pal}, SQM is absolutely stable only if the energy per baryon is less than $930.4~MeV$ \cite{Witten,Bodmer}, relative to neutron at zero external pressure. Based on this criterion, the ranges of the parameters $\beta_{\mu}$, $B_{as}$, $B_{0}$, and $\mu_{0}$ are determined by ensuring consistency with the stability condition. For fixed values of $B_{as}$ and $B_{0}$, the upper limit of $\beta_{\mu}$ is estimated for 2-flavour quark matter and the lower limit is estimated for 3-flavour quark matter. Accordingly, we select an intermediate range of parameters that is well suited for describing SQM in the present analysis. However, if $B_{as}$ and $B_{0}$ are changed, the limit of $\beta_{\mu}$ will change, so far the stability of the SQM is concerned. Following this notion, we constrain the choice of $B(\mu)$, evaluated from Eq.~(\ref{eq2}), for different values of chemical potential $(\mu)$. Now, from the Bodmer-Witten conjecture \cite{Witten,Bodmer} and the work of Pal and Chaudhuri \cite{Pal}, we assume $B_{0}=100~MeV/fm^{3}$, $B_{as}=30~MeV/fm^{3}$, $\mu_{0}=1000~MeV$ and $\beta_{\mu}=0.4-1.10$. The resulting $E_{B}-\mu$ plot is illustrated in Fig.~(\ref{fig1}). 
Following Fig.~(\ref{fig1}), we note that for different choices of $\beta_{\mu}$, we may obtain critical values of chemical potential $(\mu_{critical})$ where the $E_{B}-\mu$ plot intersects the $930.4~MeV$ line. The concerned results are tabulated in Table~(\ref{tab1}). It must be noted that, all the values of $\mu$ below $\mu_{critical}$ are allowed to ensure a stable strange matter, for which $E_{B}<930.4~MeV$ at zero external pressure \cite{Madsen}.  
\begin{table}[h]
	\centering
	\caption{Value of $\mu_{critical}$ for different values of $\beta_{\mu}$ for which $E_{B}=930.4~MeV$. \label{tab1}}
	\begin{tabular}{cc}
		\hline
		$\beta_{\mu}$ & $\mu_{critical}~(MeV)$\\
		\hline
		0.4 & 567.75 \\
		0.6 & 463.56 \\
		0.8 & 401.46 \\
		1.10 & 342.36 \\ \hline
	\end{tabular}
\end{table}
Moreover, in presence of pressure, inside the star, the energy per baryon may be greater than $930.4~MeV$ and strange matter may be stable in this situation too. According to Bodmer \cite{Bodmer} and Witten \cite{Witten} hypothesis, as we are dealing with the bare strange star modeling, we consider the value of binding energy per baryon below the value $930.4~MeV$ at the surface of the star. From Table~(\ref{tab1}), it is evident that within the stability window, the increase in control parameter $(\beta_{\mu})$ results in the increase of the range of chemical potential $(\mu)$. However, this feature may be associated with the particular choice of form for $B(\mu)$ in Eq.~(\ref{eq2}). 

Within this parameter space, we have considered the upper and lower bounds of $\beta_{\mu}$, i.e., $\beta_{\mu}=0.4$, and $\beta_{\mu}=1.10$ and varied $\mu$ within its stable range. The corresponding $B(\mu)$ values are listed in Tables~\ref{tab2a} and \ref{tab2b}. 
\begin{table}[h!]
	\centering
	\caption{Values of $B(\mu)$ for $\beta_{\mu}=0.4$ and different choices of chemical potential $(\mu).$ \label{tab2a}}
	\begin{tabular}{cc}
		\hline
		$\mu~(MeV)$ & $B(\mu)~(MeV/fm^{3})$\\
		\hline
		567.75 & 91.53 \\
		600 & 90.61 \\
		900 & 80.63 \\
		1200 & 69.35 \\
		1500 & 58.46 \\
		1526.84 & 57.55\\\hline
	\end{tabular}
\end{table}
\begin{table}[h!]
	\centering
	\caption{Values of $B(\mu)$ for $\beta_{\mu}=1.10$ and different choices of chemical potential $(\mu).$ \label{tab2b}}
	\begin{tabular}{cc}
		\hline
		$\mu~(MeV)$ & $B(\mu)~(MeV/fm^{3})$\\
		\hline
		342.36 & 91.53 \\
		350 & 91.17 \\
		400 & 88.70 \\
		600 & 77.11\\
		800 & 64.62 \\
		920.72 & 57.55 \\ \hline
	\end{tabular}
\end{table}
Interestingly, for a particular $\beta_{\mu}$, with increasing $\mu$, the system transitions to a state of higher baryon density. In this highly dense situation, the quarks are less confined. The system reaches for a deconfined phase of quarks where the vacuum pressure, i.e., difference between the  perturbative and non-perturbative vacua becomes less significant. Hence, with increasing $\mu$, $B(\mu)$ decreases.    
\section{Maximum mass-radius from TOV equations} \label{sec4} 
\begin{figure}[h!]
	\centering
	\includegraphics[width=8cm]{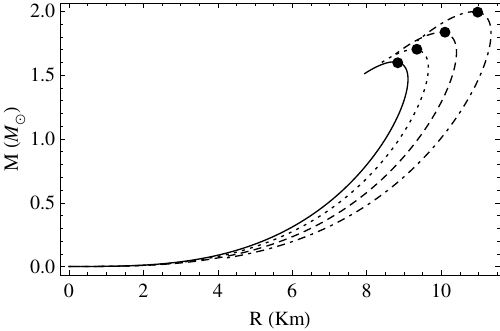}
	\caption{Mass-radius plot for $\beta=0.4$. Here, the solid, dotted, dashed and dotdashed lines represent, respectively, $B({\mu})=90.61,~80.63,~69.35$  and $58.46~\mathrm{MeV/fm^{3}}$.}
	\label{fig2a}
\end{figure}

\begin{figure}[h!]
	\centering
	\includegraphics[width=8cm]{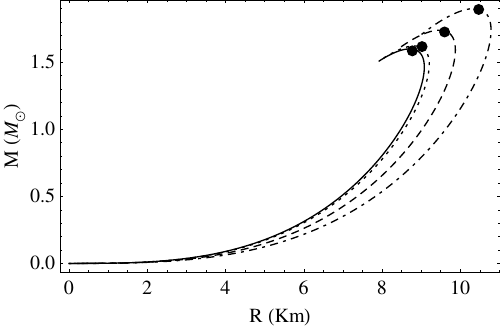}
	\caption{Mass-radius plot for $\beta=1.10$. Here, the solid, dotted, dashed and dotdashed lines represent, respectively, $B({\mu})=91.17,~88.70,~77.11,$  and $64.62~\mathrm{MeV/fm^{3}}$.}
	\label{fig2b}
\end{figure}

To derive the Tolman-Oppenheimer-Volkoff equation \cite{Tolman,Oppenheimer}, we begin with the following compact form of Einstein field equations:
\begin{equation}
	G_{\mu\nu}=R_{\mu\nu}-\frac{1}{2}g_{\mu\nu}R=8\pi\,T_{\mu\nu}, \label{eq7a}
\end{equation}
where, $G_{\mu\nu}$= Einstein tensor, $R_{\mu\nu}$= Ricci tensor, $g_{\mu\nu}$= fundamental metric tensor, $R$= Ricci scalar, and $T_{\mu\nu}$= energy-momentum tensor of the matter sector. Now, the Bianchi identity of GR, i.e., $\nabla^{\mu}G_{\mu\nu}=0$, leads to the conservation of energy-momentum tensor, expressed as $\nabla^{\mu}T_{\mu\nu}=0$. To derive the explicit form of $\nabla^{\mu}T_{\mu\nu}$, we consider the following:
\begin{itemize}
	\item a static, spherically symmetric space-time in the form:
	\begin{equation}
		ds^{2}=-e^{2\nu(r)}dt^{2}+e^{2\lambda(r)}dr^{2}+r^{2}\left(d\theta^{2}+sin^{2}\theta\,d\phi^{2}\right), \label{eq7b}
	\end{equation}
	where, $\nu(r)$ and $\lambda(r)$ are the metric potentials that solely depend on $r$. 
	\item The energy-momentum tensor for an isotropic matter distribution, {\it viz.} $T_{\mu\nu}=\text{diag}(-\rho,p,p,p)$, where $\rho$ and $p$, respectively, represent the matter variable, i.e., the energy density and pressure.  
	\item We match the interior space-time with the exterior Schwarzschild metric expressed as: 
	\begin{align}
		ds^{2}&=-\left(1-\frac{2m}{r}\right)dt^{2}+\left(1-\frac{2m}{r}\right)^{-1}dr^{2}+r^{2}\nonumber\\&\quad\left(d\theta^{2}+sin^{2}\theta\,d\phi^{2}\right), \label{eq7c}
	\end{align}
	where $m$ is the mass contained within a sphere of radius r. 
\end{itemize} 
Now, in light of the above, the Biachi identity takes the form:
\begin{equation}
	\nabla^{\mu}T_{\mu\nu}=\frac{dp}{dr}+(\rho+p)\frac{d\nu}{dr}. \label{eq7d}
\end{equation}
To simplify Eq.~(\ref{eq7d}), we consider the $G_{tt}$ component of the field equation and obtain $\frac{d\nu}{dr}=\frac{4\pi\,r^{3}p+m(r)}{r(r-2m(r))}$. Substituting this expression back into Eq.~(\ref{eq7d}), we obtain:
\begin{equation}
	\frac{dp}{dr}=-\Bigg(\frac{\rho+p}{r^{2}}\Bigg)\Bigg(m+4\pi r^{3}p\Bigg)\Bigg(1-\frac{2m}{r}\Bigg)^{-1}, \label{eq7}
\end{equation}
and from $G_{rr}$ component we obtain, 
\begin{equation}
	\frac{dm}{dr}=4\pi r^{2}\rho. \label{eq8} 
\end{equation}
Eqs.~(\ref{eq7}) and (\ref{eq8}) are termed the TOV equations. Notably, to present a neat expression, we have dropped the functional dependencies, i.e., $m(r)=m$, $p(r)=p$, $\rho(r)=\rho$. Simultaneous solution of Eqs.~(\ref{eq7}) and (\ref{eq8}) using the EoS expressed in Eq.~(\ref{eq6}) and $B(\mu)$ listed in Tables~\ref{tab2a} and \ref{tab2b}, along with the boundary condition, $m(0)=0$, $p(0)=\text{central pressure}$ and $\rho(0)=\text{central density}$, yield the maximum mass and the associated radius in the present formalism. Figs.~(\ref{fig2a}) and (\ref{fig2b}) illustrate the mass-radius plot. 
From the Fig.~(\ref{fig2a}) and (\ref{fig2b}), it is noted that with decreasing $B(\mu)$, the maximum mass and radius increases, which may be explained in the following way: with decreasing $B(\mu)$, the difference of vacuum energies decreases, thereby increasing the pressure component. With increasing pressure, the EoS becomes stiffer which leads to increased maximum mass and the corresponding radius. The concerned results are tabulated in Tables~(\ref{tab3a}) and (\ref{tab3b}). 
\begin{table}[h!]
	\centering
	\caption{Tabulation of maximum mass-radius for $\beta=0.4$\label{tab3a}}
	\begin{tabular}{ccc}
		\hline
		$B(\mu)~(\mathrm{MeV/fm^{3}]}$ & $M~(\mathrm{M_{\odot}})$ & $R~(\mathrm{Km})$\\
		\hline
		90.61 &	1.60 & 8.73 \\
		80.63 &	1.70 & 9.26 \\ 
		69.35 &	1.83 & 9.98	\\
		58.46 &	1.99 & 10.88 \\	\hline		
	\end{tabular}
\end{table}
\begin{table}[h!]
	\centering
	\caption{Tabulation of maximum mass-radius for $\beta=1.10$\label{tab3b}}
	\begin{tabular}{ccc}
		\hline
		$B(\mu)~(\mathrm{MeV/fm^{3}]}$ & $M~(\mathrm{M_{\odot}})$ & $R~(\mathrm{Km})$\\
		\hline
		91.17 &	1.59 & 8.71 \\
		88.70 &	1.62 & 8.83 \\
		77.11 &	1.74 & 9.47 \\
		64.62 &	1.90 & 10.35 \\		\hline				
	\end{tabular}
\end{table}
From Tables~(\ref{tab3a}) and \ref{tab3b}, it is evident that the maximum mass achieved in this parameter space is $\approx2~M_{\odot}$. Therefore, this model is suitable to describe the low mass gap of compact stars. To assess the physical acceptability of the model, we have predicted the radii of some low mass compact objects from the mass-radius relation. Notably, for this purpose, we have fixed $\beta_{\mu}$ at 0.6. The results are tabulated in Table~(\ref{tab4}). 
\begin{table*}[ht]
	\centering
	\caption{Radius prediction of low mass compact objects\label{tab4}}
	\begin{tabular}{cccccc}
		\hline
		\multirow{3}{*}{Compact Objects} & Measured mass & Measured radius & $B(\mu)$ & $\mu$ & Predicted radius \\
		& from observation & from observation & $(MeV/fm^{3})$ & $(MeV)$ & from model \\
		& $(M_{\odot})$ & $(Km)$ &  &  & $(Km)$\\
		\hline
		Her X-1 \cite{Abubekerov} & $0.85\pm0.15$ & $8.1\pm0.41$ & 91.00 & 478.92 & 8.13 \\
		LMC X-4 \cite{Rawls} & $1.04\pm0.09$ & $8.301\pm0.2$ & 91.30 & 470.31 & 8.50 \\
		SMC X-4 \cite{Rawls} & $1.29\pm0.05$ & $8.831\pm0.09$ & 91.45 & 465.96 & 8.95 \\
		Cen X-3 \cite{Rawls} & $1.49\pm0.08$ & $9.178\pm0.13$ & 89.00 & 533.78 & 9.19 \\
		4U 1820-30 \cite{Guver} & $1.58\pm0.06$ & $9.1\pm0.4$ & 89.00 & 533.78 & 9.10 \\ \hline
	\end{tabular}
\end{table*}
From Table~(\ref{tab4}), it is noted that using the present formalism, we can obtain the radii of the low mass compact objects with great accuracy. 
\section{Tidal deformability and Love number}\label{sec5} The Tidal Love Number (TLN), commonly represented as $k_{2}$, quantifies the extent to which a compact astrophysical object deforms in response to an external gravitational tidal field. It characterises the induced quadrupole moment relative to the perturbing tidal potential. In parallel, the dimensionless tidal deformability parameter $(\Lambda)$ provides a measure of the star's susceptibility to such deformations, reflecting the degree of distortion experienced at the surface. Now, $k_{2}$ and $\Lambda$ are interconnected through a mathematical relation that links the internal structure of the star to its response under tidal forces as:
\begin{equation}
	k_{2}=\frac{3}{2}\Lambda\Big(\frac{M}{R}\Big)^{5}, \label{eq9}
\end{equation}
where, $M$ and $R$ represent the mass and radius of the object. To begin with, we consider the linear perturbations owing to an external quadrupolar tidal field \cite{Thorne} and the space-time metric is written as:
\begin{equation}
	g_{\mu\nu}=g_{\mu\nu}^{0}+h_{\mu\nu}, \label{eq10}
\end{equation}
where, the linearised metric perturbations are characterised by $h_{\mu\nu}$. In the Regge-Wheeler static, even parity $(\ell=2)$ framework \cite{Regge}, $h_{\mu\nu}$ is expressed as \cite{Hinderer}:
\begin{align}
	h_{\mu\nu}=&diag\Big[-e^{2\nu(r)}H_{0}(r),e^{2\lambda(r)}H_{2}(r),r^{2}K(r),\nonumber\\&\quad\,r^{2}sin^{2}\theta K(r)\Big]Y_{2m}(\theta,\phi). \label{eq11}
\end{align}
Now, following the detailed steps described in the work of T. Hinderer \cite{Hinderer}, we may study the tidal deformation and the associated love numbers in the present context. We begin with the linearised Einstein field equations in the form:
\begin{equation}
	\delta G_{\mu\nu} = 8\pi\delta T_{\mu\nu}.\label{eq13}
\end{equation}
Substituting the perturbed metric given in Eq.~(\ref{eq11}) together with the corresponding perturbations of the stress-energy tensor into Eq.~(\ref{eq13}) we derive a set of coupled differential equations for the metric perturbation functions. For a perfect fluid configuration, the non-vanishing perturbations of the stress-energy tensor are given by
\begin{equation}
	\delta T^{0}_{0}=-\delta \rho
	=-\left(\frac{dp}{d\rho}\right)^{-1}\delta p,
	\qquad
	\delta T^{i}_{i}=\delta p. \label{eq13a}	
\end{equation}
The difference between the angular components of the perturbed Einstein equations, i.e., $\delta G^{\theta}_{\theta}-\delta G^{\phi}_{\phi}=0$ leads to the equality $H_{2}(r)=H_{0}(r)\equiv H(r).$ In addition, the mixed component $\delta G^{r}_{\theta}=0$ provides a relation involving the perturbation function $(K(r))$ and the radial function $(H(r))$. By combining the angular field equations, the pressure perturbation $(\delta p)$ can be eliminated. Subsequently, subtracting the rr-component from the tt-component of the perturbed Einstein equations yields the following decoupled second-order differential equation governing the radial metric perturbation $(H(r))$ for the static even parity perturbation $(\ell=2)$:
\begin{align}
	H^{\prime\prime}(r)+A_{1}H^{\prime}(r)+A_{2}H(r)=0, 
	\label{eq13b}
\end{align}
where 
\begin{equation}
	A_{1}=\frac{1+e^{2\lambda}(1+4\pi\,r^2p-4\pi r^2\rho)}{r}, \label{eq13b1}
\end{equation}
and,
\begin{align}
	A_{2}=&-\frac{1+4e^{2\lambda}+e^{4\lambda}}{r^2}+4\pi\,e^{2\lambda}(p+\rho)+\frac{4\pi\,e^{2\lambda}(p+\rho)}{c^2}\nonumber\\&\quad
	-e^{4\lambda}r^2\{8\pi p\}^{2}-2e^{2\lambda}\{8\pi p(e^{2\lambda}-3)-8\pi\rho\}, \label{eq13b2}
\end{align}
where $c^{2}=dp/d\rho$ denotes the squared speed of sound. In the exterior region of a compact object, the space-time is governed by the Schwarzschild solution. By imposing the vacuum conditions $\rho=0$ and $p=0$, together with $e^{2\lambda}=\left(1-\frac{2M}{r}\right)^{-1}$, the master Eq.~(\ref{eq13b}) reduces to the following second-order differential equation for $H(r)$:
\begin{equation}
	H^{\prime\prime}(r)
	+\frac{2(M-r)}{r(2M-r)}H^{\prime}(r)
	-\frac{2\left(2M^{2}-6Mr+3r^{2}\right)}
	{r^{2}(r-2M)^{2}}H(r)=0.
	\label{eq13b3}
\end{equation}
The general solution of Eq.~(\ref{eq13b3}) may be expressed as
\begin{align}
	H(r)=&c_{1}\Bigg(\frac{r}{M}\Bigg)^{2}
	\Bigg(1-\frac{2M}{r}\Bigg)
	\Bigg[
	-\frac{M(M-r)(2M^{2}+6Mr-3r^{2})}
	{r^{2}(2M-r)^{2}}\nonumber\\&\quad
	+\frac{3}{2}\ln\Bigg(\frac{r}{r-2M}\Bigg)
	\Bigg]
	+3c_{2}\Bigg(\frac{r}{M}\Bigg)^{2}
	\Bigg(1-\frac{2M}{r}\Bigg).
	\label{eq13b4}
\end{align}

The asymptotic nature of Eq.~(\ref{eq13b4}) takes the following form:
\begin{equation}
	H(r)
	=
	\frac{8}{5}c_{1}\left(\frac{M}{r}\right)^{3}
	+\mathcal{O}\left(\left(\frac{M}{r}\right)^{4}\right)
	+3c_{2}\left(\frac{r}{M}\right)^{2}
	+\mathcal{O}\left(\frac{M}{r}\right).
	\label{eq13b5}
\end{equation}
Matching this asymptotic expansion with the far-field metric limit associated with the external tidal field and induced quadrupole moment yields,
\begin{equation}
	c_{1}=\frac{15}{8}\frac{k\mathcal{E}}{M^{3}},
	\qquad
	c_{2}=\frac{1}{3}M^{2}\mathcal{E},
	\label{eq13i}
\end{equation}
where $\mathcal{E}$ denotes the external quadrupolar tidal field and $k$ is the tidal deformability parameter. Finally, imposing continuity of the perturbation function and its derivative at the stellar surface $(r=R)$, the dimensionless quadrupolar tidal Love number $(k_{2})$ is obtained as,
\begin{align}
	k_{2}&=\frac{8u^{5}}{5}(1-2u)^{2}[2+2u(g-1)-g]\nonumber\\&\quad\times\Bigg[2u[6-3g+3u(5g-8)]+4u^{3}[13-11g\nonumber\\&\quad+u(3g-2)+2u^{2}(1+g)]+3(1-2u)^{2}[2-g\nonumber\\&\quad+2u(g-1)ln(1-2u)]\Bigg]^{-1}, \label{eq12}
\end{align}
where, $u=\frac{M}{R}$ represents the compactness and $g=\frac{RH'(R))}{H(R)}$ is evaluated from the numerical integration of the interior perturbation equation. In the present paper, we have obtained the tidal parameters with the help of Eq.~(\ref{eq6}). To avoid the mathematical complexities, we have illustrated the results graphically. 
\begin{figure}[h!]
	\centering
	\includegraphics[width=8cm]{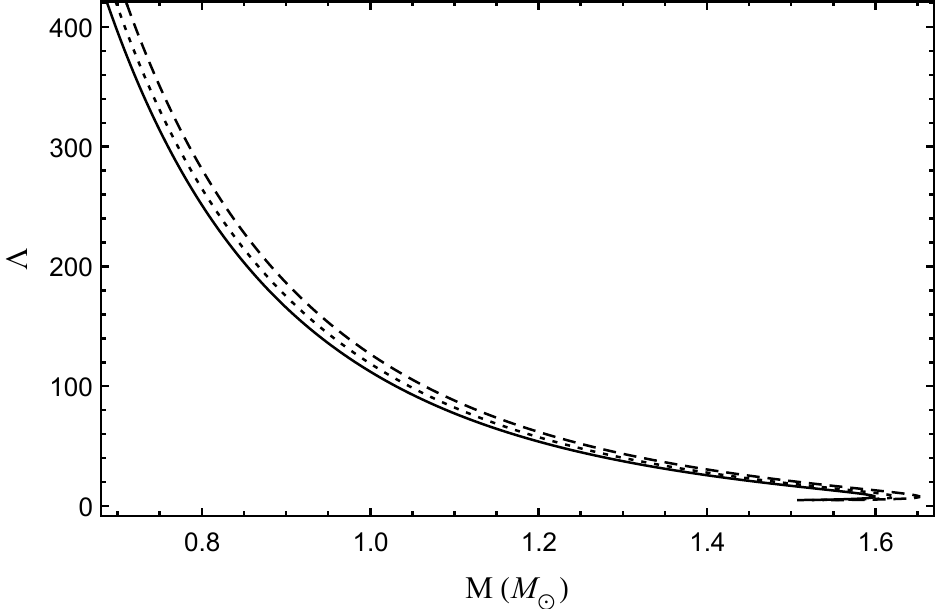}
	\caption{Variation of tidal deformability $(\Lambda)$ with maximum mass $(M_{\odot})$. Here, the solid, dotted and dashed lines represent, respectively, $B({\mu})=91.14,~88.38$ and $85.37~\mathrm{MeV/fm^{3}}$.}
	\label{fig3}
\end{figure}
From Fig.~(\ref{fig3}), we note two interesting features, {\it viz.}, (i) for a particular $B(\mu)$, the tidal deformability $(\Lambda)$ decreases with increasing mass of the stellar configuration, and (ii) for a particular mass, $\Lambda$ increases with decreasing $B(\mu)$. These results can be attributed to the fact that a smaller bag parameter implies higher pressure and lower compressibility at a given central density, i.e. a stiffer EoS. From Tables~(\ref{tab3a}) and (\ref{tab3b}), it is evident that a stiffer EoS points toward larger values of mass and radius. Therefore, in the first case, it becomes a difficult task to deform the object with comparatively high mass value. Further, from Eq.~(\ref{eq9}), we note that $\Lambda\propto\Big(\frac{M}{R}\Big)^{-5}$ and for a fixed mass, decreasing $B(\mu)$ results in larger radius and lower compactness. Hence, with decreasing $B(\mu)$, even a small decrease in compactness leads to large increment in $\Lambda$. 
\begin{figure}[h!]
	\centering
	\includegraphics[width=8cm]{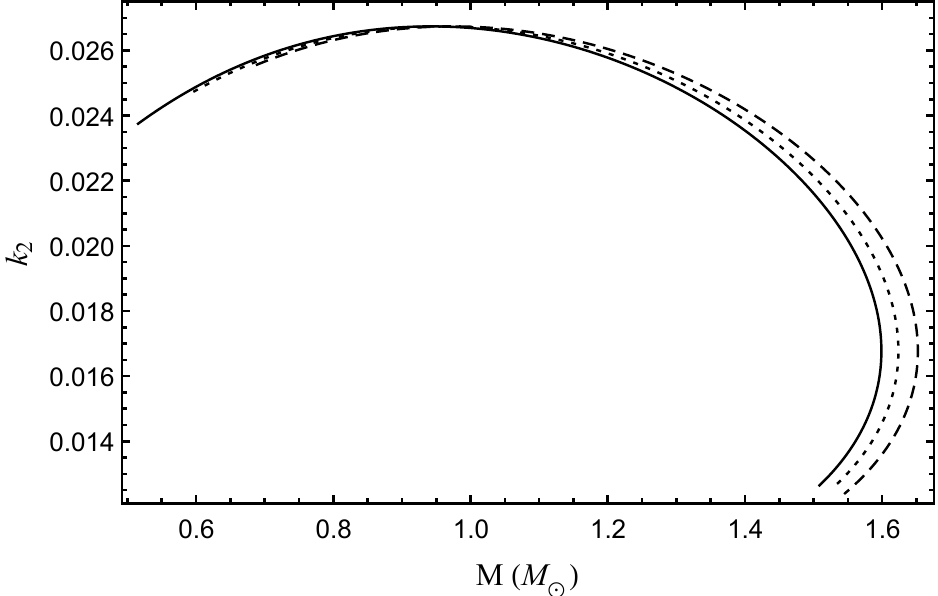}
	\caption{Variation of tidal love number $(k_{2})$ with maximum mass $(M_{\odot})$. Here, the solid, dotted and dashed lines represent, respectively, $B({\mu})=91.14,~88.38$ and $85.37~\mathrm{MeV/fm^{3}}$. }
	\label{fig4}
\end{figure}
Notably, Fig.~(\ref{fig4}) illustrates that for a particular $B(\mu)$, love number $k_{2}$ increases with increasing mass, reaches a maximum value and then converges to the maximum mass point associated with the particular choice of $B(\mu)$. On the other hand, for a particular mass, $k_{2}$ increases with decreasing $B(\mu)$. To describe the nature of this variation, we study of the nature of $\Lambda$ and $k_{2}$ vs. compactness plots, as shown in Figs.~(\ref{fig5}) and (\ref{fig6}), respectively.   
\begin{figure}[h!]
	\centering
	\includegraphics[width=8cm]{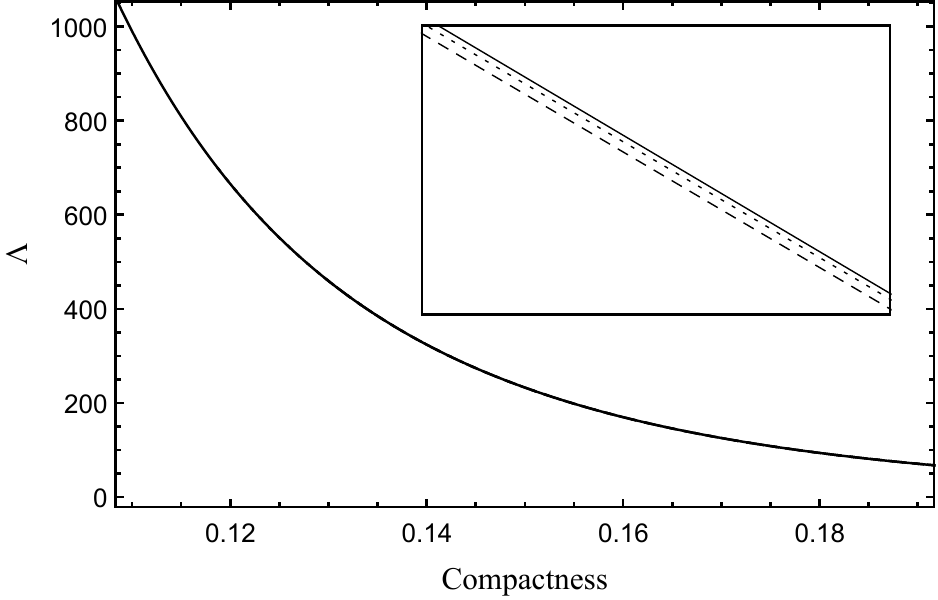}
	\caption{Variation of tidal deformability $(\Lambda)$ with compactness. Here, the solid, dotted and dashed lines represent, respectively, $B({\mu})=91.14,~88.38,~85.37$. and $82.17~\mathrm{MeV/fm^{3}}$.}
	\label{fig5}
\end{figure}
\begin{figure}[h!]
	\centering
	\includegraphics[width=8cm]{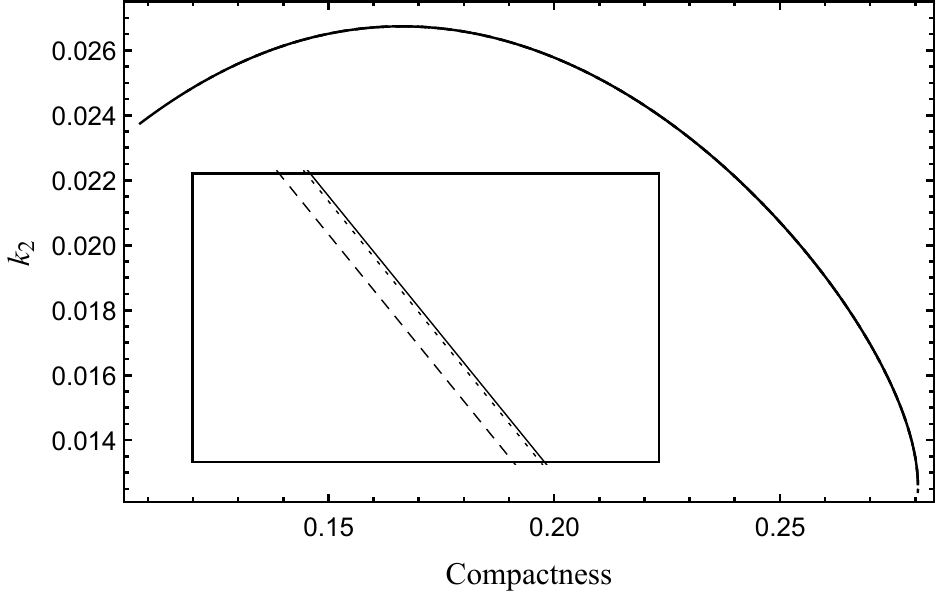}
	\caption{Variation of tidal love number $(k_{2})$ with compactness. Here, the solid, dotted and dashed lines represent, respectively, $B({\mu})=91.14,~88.38$ and $85.37~\mathrm{MeV/fm^{3}}$.}
	\label{fig6}
\end{figure}
From Fig.~(\ref{fig5}), we note that $\Lambda$ decreases with increasing compactness which is an expected result in the present scenario. Further, from Fig.~(\ref{fig6}), we note that with increasing compactness, $k_{2}$ first increases, reaches a maximum point and then decreases, exhibiting a similar nature as shown in Fig.~(\ref{fig4}). This may be explained in the following way: when the compactness is low, the object's gravitational binding is less and the external tidal forces induce a mass quadrupole moment, since it does not resist the tidal deformation very well. Hence, we obtain a large $\Lambda$. Now, with increasing compactness, the self-gravity becomes stronger. This enhances the induced quadrupole moment because the structure is still deformable, but now it is coherent. So, quadrupole moment increases, and as a results, from Eq.~(\ref{eq9}), $k_{2}$ increases. Up to certain extent, the love number increases and becomes optimum. Further, as compactness increases more, the object's resistance to deformation grows. Now, the internal pressure becomes high and the space-time curvature becomes intense, implying that the domination of relativistic effects. This effectively means that the same tidal field induces a smaller quadrupole moment now, which in turn yields a smaller love number $k_{2}$. Therefore, with further increment in compactness, $k_{2}$ decreases. Moreover, from Fig.~(\ref{fig4}), we have found that with increasing mass, the value of $k_{2}$ approaches the maximum mass points. Additionally, From Figs.~(\ref{fig5}) and (\ref{fig6}), it is evident that for a particular compactness, $\Lambda$ and $k_{2}$ decrease with decreasing $B(\mu)$. From Eq.~(\ref{eq9}), we note that for a particular compactness, $k_{2}\propto\Lambda$. In this context, with decreasing $B(\mu)$, the pressure increases which resists the tidal deformation. Hence, $\Lambda$ and $k_{2}$ decrease with decreasing $B(\mu)$ for a particular compactness.   

Now, the physical inference of the above analysis is incomplete without checking the applicability of this theoretical framework of tidal parameters, on different class of compact stars. Prior to the successful detection of gravitational waves, the bounds of the tidal parameters remained elusive. However, the binary neutron star merger event GW170817, analysed by Abbott et al. \cite{Abbott}, placed constraints on the tidal deformability parameter $\Lambda$. According to the findings reported in the articles \cite{Abbott,Bauswein}, in case of low-spin prior, the tidal deformability $(\Lambda)$ for a NS with mass $1.4~M_{\odot}$ must not exceed a value of 800. Moreover, we have noted that the evaluation of $g$, in Eq.~(\ref{eq12}), depends on the squared adiabatic sound velocity, and hence in a physically realistic system, the solution may be modified due to a phase transition emerging from the presence of finite surface density. When more than one conserved quantity, such as electric charge and baryon number, participates in a phase transition, the requirement of global charge neutrality, together with the Gibbs phase equilibrium conditions, can result in a smooth variation of pressure with energy density. Consequently, even a first-order phase transition need not produce a finite discontinuity in the energy density. The effects of a discontinuous energy density profile on neutron star structure were investigated by Damour and Nagar \cite{Damour}. Their analysis demonstrated that a sufficiently large jump in the energy density across the transition can lead to a substantial modification of the quadrupolar tidal Love number, $k_{2}$. Now, the sound velocity in the presence of finite surface density can be expressed as:
\begin{equation}
	\frac{d\rho}{dp}=\frac{d\rho}{dp}\Bigg{|}_{p\neq\,p_{\text{d}}}+\Delta\,\rho\,\delta(p-p_{\text{d}}), \label{eq12a}
\end{equation}
where the pressure at the discontinuity is denoted by $p_{\text{d}}$, and $\Delta\,\rho=\rho(p_{\text{d}}+0)-\rho(p_{\text{d}}-0)$ denotes the jump due to a finite surface density correction. Hence, the finite surface density corrected $g$ is expressed as:
\begin{equation}
	g(r_{\text{d}}+\epsilon)=	g(r_{\text{d}}-\epsilon)-\frac{\rho(r_{\text{d}}+\epsilon)-\rho(r_{\text{d}}-\epsilon)}{m(r_{\text{d}})/\bar{\rho}}, \label{eq12b}
\end{equation}
where $\epsilon\rightarrow0$, $\bar{\rho}=4\pi\,r_{\text{d}}^{3}$ represents the interior mean density, and $r_{\text{d}}$ is the point of discontinuity. To attest these predictions, we have extracted the values of $\Lambda$ and $k_{2}$, in both their uncorrected form and after applying the finite surface density correction, for different low mass compact stars, and presented them in Table~\ref{tab5}. 
\begin{table*}[t!]
	\centering
	\caption{Tabulation of tidal love number and tidal deformability of low mass compact objects\label{tab5}}
	\begin{tabular}{ccccccc}
		\hline
		Compact objects & Measured mass & $\mu$ & $k_{2}$ & $\Lambda$ & $(k_{2})_{cor}$ & $(\Lambda)_{cor}$\\
		& $(M_{\odot})$ & $(MeV)$ & & & & \\
		\hline
		\multirow{4}{*}{Her X-1} & \multirow{4}{*}{$0.85\pm0.15$} & 475 & 0.0266128 & 203.564 & 0.237959 & 1820.18\\ 
		&& 550 & 0.0265821 & 212.712 & 0.241028 & 1928.72 \\
		&& 625 & 0.0265233 & 229.104 & 0.246218 & 2126.78 \\
		\hline
		\multirow{4}{*}{SAXJ 1808.4-3658} & \multirow{4}{*}{$0.9\pm0.3$} & 475 & 0.0267136 & 165.213 & 0.223456 & 1381.99 \\
		&& 550 & 0.0266907 & 176.143 & 0.227894 & 1503.97 \\
		&& 625 & 0.0266626 & 187.056 & 0.232069 & 1628.11 \\
		\hline
		\multirow{4}{*}{LMC X-4} & \multirow{4}{*}{$1.04\pm0.09$} & 475 & 0.0266034 & 96.8826 & 0.18713 & 681.48 \\
		&& 550 & 0.0266405 & 102.257 & 0.190739 & 732.133 \\
		&& 625 & 0.0266779 & 109.324 & 0.19523 & 800.041 \\
		\hline
		\multirow{4}{*}{SMC X-4} & \multirow{4}{*}{$1.29\pm0.05$} & 475 & 0.0249743 & 38.8763 & 0.129423 & 201.467\\
		&& 550 & 0.0251635 & 41.5911 & 0.13343 & 220.538 \\
		&& 625 & 0.0253684 & 44.975 & 0.138132 & 244.89 \\
		\hline
		\multirow{4}{*}{Cen X-3} & \multirow{4}{*}{$1.49\pm0.08$} & 475 & 0.0218463 & 17.5289 & 0.0861267 & 69.1059 \\
		&& 550 & 0.0223449 & 19.4277 & 0.0912717 & 79.3561 \\
		&& 625 & 0.0228219 & 21.5659 & 0.0966364 & 91.318 \\
		\hline
		\multirow{4}{*}{4U 1820-30} & \multirow{4}{*}{$1.58\pm0.06$} & 475 & 0.0188717 & 10.4072 & 0.0622332 & 34.3198 \\
		&& 550 & 0.019958 & 12.4034 & 0.0698645 & 43.4189 \\
		&& 625 & 0.020837 & 14.4621 & 0.0768849 & 53.3627 \\
		\hline
	\end{tabular}
\end{table*}
In Table~\ref{tab5}, we have also shown the impact of chemical potential on $\Lambda$ and $k_{2}$. It is to be noted that Table~\ref{tab5} maintains the physical justification given above. Moreover, within this parameter space, the bound $\Lambda<800$ \cite{Abbott} is well satisfied, which points towards a viable framework to study low mass compact objects.  
\section{Conclusions}\label{sec6}
\begin{figure}[h!]
	\centering
	\includegraphics[width=8cm]{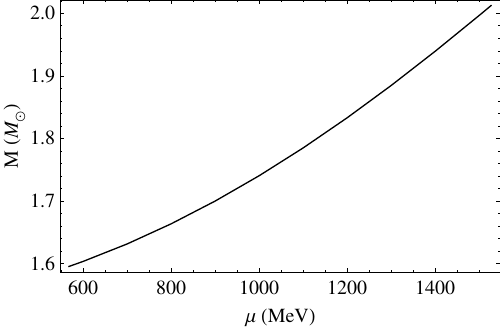}
	\caption{Variation maximum mass with chemical potential $(\mu)$.}
	\label{fig7}
\end{figure}

In the present analysis, we have studied the quark matter medium effects through the incorporation of chemical potential dependent bag parameter in the context of MIT bag model EoS. We have adhered to the formalism put forward by Pal and Chaudhuri \cite{Pal} where the EoS has been formulated using the Euler approach. Notably, we use the chemical potential dependency to ensure consistency of the present framework with the grand canonical ensemble. However, using such an approach requires careful supervision of the parameter space. Considering zero external pressure relative to neutron, we have computed the energy per baryon $(E_{B})$, and studied the absolute stability of the SQM. Following the Bodmer-Witten \cite{Witten,Bodmer} hypothesis, we have constrained the parameter space in the limit $E_{B}<930.4~\mathrm{MeV}$, which yields the critical bounds for the chemical potential $(\mu)$ as tabulated in Table~\ref{tab1}. Using a Gaussian bag parametrisation along with the upper and lower bounds of the control parameter $\beta_{\mu}$ \cite{Pal}, we have evaluated and tabulated the value of bag parameter $B(\mu)$, in Tables~\ref{tab2a} and \ref{tab2b}. From Tables~\ref{tab2a} and \ref{tab2b}, we note that with increasing $\mu$, $B(\mu)$ decreases, which may be explained in the following way: with increasing chemical potential, the system gets densely packed, and the energy cost of adding another particle increases. In this high density environment, the QCD vacuum difference decreases since the quark de-confinement becomes easier. As a result, the bag parameter decreases. Further, we have studied the physical aspect of the present framework through two different methodologies, i.e., (i) the maximum mass-radius relation by solving the TOV equations \cite{Tolman,Oppenheimer}, and (ii) the tidal deformability and tidal love number. Using the results of Tables~\ref{tab2a} and \ref{tab2b}, we have solved the TOV equations, and obtained the maximum mass and the associated radius. The results are tabulated in Tables~\ref{tab3a} and \ref{tab3b}. Notably, within this parameter space, the maximum achievable stellar mass is $\approx2~M_{\odot}$. Therefore, the present formalism is well-suited to describe the lower mass-gap of compact stars. Notably, with decreasing $B(\mu)$, the maximum mass increases owing to stiffening of EoS. We further use the mass-radius relation to predict the radii of some SS candidates, in the lower mass-gap region, for different choices of $\mu$, and associated $B(\mu)$, in Table~\ref{tab4}. We have noted that in the present framework, we can predict the radii of different SS candidates with greater accuracy. In the conventional MIT bag model with a constant bag $(B)$, the maximum mass-radius of compact stars are fixed once a particular value of $B$ is chosen in EoS. Consequently, obtaining different maximum mass-radius requires changing the bag constant itself, which introduces significant arbitrariness and makes systematic exploration cumbersome. In contrast, a chemical potential dependent bag model yields a smooth and continuous variation of the maximum mass and associated radius with the chemical potential $\mu$. For a given compact star, the chemical potential associated with a particular mass can be inferred from Fig.~\ref{fig7}. This feature enables a more detailed and physically transparent probe of the stellar interior. Moreover, the continuous correlation between $\mu$ and stellar mass allows different mass configurations to be generated simply by tuning the chemical potential, without the need to modify the underlying model parameters. This represents a significant advantage over the constant bag model in studying the internal structure and properties of compact stars which is evident from Fig.~\ref{fig7}. 
Next, we have studied the impact of chemical potential on tidal deformability and TLN. These tidal parameters are formulated on the basis of the calculations demonstrated by Hinderer \cite{Hinderer}, in the context of static, even parity perturbations. For the present parametric choices, and the mass-radius results, we have plotted the variations of these tidal parameters in Figures~\ref{fig3}-\ref{fig6}. Moreover, in Table~\ref{tab5}, we have shown the numerical variations of tidal parameters along with their respective finite surface density corrected values with chemical potential $(\mu)$. Notably, with increasing $\mu$, the bag parameter $B(\mu)$ decreases. This decrease in $B(\mu)$ results in a stiffer EoS, thereby increasing the mass-radius of the system. Now, from the definition, $\Lambda\propto \Big(\frac{M}{R}\Big)^{-5}$, and for a particular mass, a decreasing $B(\mu)$ produces a larger radius. Consequently, $\Lambda$ increases which is evident from Table~\ref{tab5}. Now, to substantiate our findings, we turn to the merger event GW170817 \cite{Abbott} and invoke the restriction, $\Lambda<800$ for a NS with a mass of $1.4~M_{\odot}$. Notably, the bound $\Lambda<800$ is well maintained in the present analysis. However, no constrained limit has been obtained for the corrected tidal parameters, to date. Hence, the finite surface density corrected values may remain well above the $\Lambda$ values. In the present model, it is noted that tidal properties have been modified compared to the conventional models. A stiffer EoS corresponds to larger radii and higher maximum masses. Conventional models with constant $B$ typically predict smaller radii and lower maximum masses for the same central density. The value of $\Lambda \propto (M/R)^{-5}$, so increasing the radius at fixed mass boosts $\Lambda$ which, in turn, increases $k_{2}$. Conventional models miss this subtle interplay and  underestimate $\Lambda$ for the same stellar mass because they have smaller radii. $k_{2}$ and $\Lambda$ are tightly coupled to radius and compactness, which are fixed differently in constant $B$ models. Accordingly, chemical potential dependence allows more massive stars to be less compact at the same mass, enhancing tidal deformability, unlike conventional models. The mass-radius-tidal deformability relation is shifted, predicting larger $\Lambda$ for a given mass and a more nuanced behaviour of $k_{2}$ as a function of compactness.

It must be noted that we have fine-tuned the model parameters within the permissible range to obtain the results. However, there are certain limitations in the present model: (i) The results related to GW170817, used here are itself depend on the assumptions of low-spin priors, quasi-circular binary inspiral etc. Hence, the obtained bound is not entirely universal for different NS EoSs. (ii) The obtained range of chemical potential in our model is constrained by the Bodmer-Witten hypothesis. Hence, a particular range of chemical potential is only permissible here if $B_{0},~B_{as}$ and $\beta_{\mu}$ are chosen. However, different choices of $B_{0},~B_{as}$ modify the value of $\beta_{\mu}$ and hence the allowed range of chemical potential. So, in a sense, this agreement between our result and that of GW170817 is not generic, rather intrinsically fine-tuned. (iii) The present analysis is restricted to static, cold, and non-rotating compact stars. In a more physically realistic setting, however, effects due to finite temperature, rotation, and magnetic fields are expected to play an important role and can significantly influence the tidal properties of compact stars. Although our model shows agreement with observed properties of GW170817 within this parameter space, suggesting the physical viability of the framework. 

Overall, the present investigation demonstrates that incorporating a chemical potential–dependent bag parameter within the MIT bag model framework provides a more realistic description of strange quark matter and its macroscopic manifestations. The obtained results successfully reproduce the observed stellar masses and tidal deformability constraints from GW170817, thereby establishing the suitability of the present formalism in modeling strange stars within the lower mass-gap $(\leq2.01~M_{\odot})$ region of compact stars. 

\section*{Acknowledgments}
DB is thankful to Department of Science and Technology (DST), Govt. of India, for providing the fellowship vide no: DST/INSPIRE Fellowship/2021/IF210761. PKC gratefully acknowledges the support from IUCAA, Pune, India for providing the Visiting Associateship programme. DB and PKC gratefully acknowledge the facilities provided by IUCAA during their visit, where this work was completed.

%\appendix

%\section{Appendixes}
%\end{verbatim}
\nocite{*}
%\section*{References}

\end{document}